\documentclass[aps,twocolumn,amssymb,amsmath,superscriptaddress]{revtex4-1}

\usepackage{bm}
\usepackage{color}
\usepackage{graphicx}
\usepackage{dcolumn}
\usepackage{graphicx}
\usepackage[colorlinks=true, letterpaper=true, pdfstartview=FitV, linkcolor=blue, citecolor=blue, urlcolor=blue]{hyperref}
\usepackage[normalem]{ulem}
\usepackage{amsmath}
\usepackage{epstopdf}
\usepackage{multirow}
\usepackage{makecell} 
\usepackage{booktabs}
\usepackage{float}
\usepackage[utf8]{inputenc}
\usepackage[T1]{fontenc}

\begin{document}

\title{Giant Magneto-Optical Effects in Two-Dimensional Flat-Band Antiferromagnets}

\author{Ping Yang}
\affiliation{Key Lab of Advanced Optoelectronic Quantum Architecture and Measurement (MOE), Beijing Key Lab of Nanophotonics and Ultrafine Optoelectronic Systems, and School of Physics,Beijing Institute of Technology, Beijing 100081, China}
\affiliation{Department of Physics, Tianjin Renai College, Tianjin 301636, China}

\author{Wanxiang Feng}
\email{wxfeng@bit.edu.cn}
\affiliation{Key Lab of Advanced Optoelectronic Quantum Architecture and Measurement (MOE), Beijing Key Lab of Nanophotonics and Ultrafine Optoelectronic Systems, and School of Physics,Beijing Institute of Technology, Beijing 100081, China}

\author{Siyuan Liu}
\affiliation{Key Lab of Advanced Optoelectronic Quantum Architecture and Measurement (MOE), Beijing Key Lab of Nanophotonics and Ultrafine Optoelectronic Systems, and School of Physics,Beijing Institute of Technology, Beijing 100081, China}

\author{Shan Guan}
\affiliation{State Key Laboratory of Semiconductor Physics and Chip Technologies, Institute of Semiconductors, Chinese Academy of Sciences, Beijing 100083, China}

\author{Liwei Wen}
\affiliation{College of Science, Henan University of Engineering, Zhengzhou 451191, China}

\author{Wei Jiang}
\affiliation{Key Lab of Advanced Optoelectronic Quantum Architecture and Measurement (MOE), Beijing Key Lab of Nanophotonics and Ultrafine Optoelectronic Systems, and School of Physics,Beijing Institute of Technology, Beijing 100081, China}
\affiliation{International Center for Quantum Materials, Beijing Institute of Technology, Zhuhai, 519000, China}

\author{Gui-Bin Liu}
\affiliation{Key Lab of Advanced Optoelectronic Quantum Architecture and Measurement (MOE), Beijing Key Lab of Nanophotonics and Ultrafine Optoelectronic Systems, and School of Physics,Beijing Institute of Technology, Beijing 100081, China}

\author{Yugui Yao}
\email{ygyao@bit.edu.cn}
\affiliation{Key Lab of Advanced Optoelectronic Quantum Architecture and Measurement (MOE), Beijing Key Lab of Nanophotonics and Ultrafine Optoelectronic Systems, and School of Physics,Beijing Institute of Technology, Beijing 100081, China}
\affiliation{International Center for Quantum Materials, Beijing Institute of Technology, Zhuhai, 519000, China}

\date{\today}

\begin{abstract}
In this work, we reveal giant magneto-optical responses in two-dimensional (2D) antiferromagnets with nearly flat electronic bands, based on first-principles calculations and group-theoretical analysis. We identify a record-large second-order magneto-optical Schäfer-Hubert (SH) effect—featuring a polarization rotation angle of $28^\circ$—in monolayer antiferromagnetic RuOCl$_2$, driven by flat-band–enhanced interband optical transitions. Both the valence and conduction bands exhibit pronounced directional flatness, giving rise to highly anisotropic optical absorption and broadband hyperbolic frequency windows spanning the entire visible spectrum. This anisotropy leads to an exceptionally strong linear dichroism (LD) reaching 50\%, far exceeding values reported in other 2D magnetic systems. Remarkably, the giant SH effect and LD appear at distinct photon energies, reflecting a momentum-direction–dependent crossover between flat and dispersive bands. Both responses are further amplified with increasing RuOCl$_2$ film thickness. Our results establish flat-band antiferromagnets as a fertile platform for realizing giant nonlinear magneto-optical effects and open new avenues for 2D opto-spintronic device applications.
\end{abstract}

\maketitle

Antiferromagnets, with vanishing net magnetization, terahertz-scale spin dynamics, and inherent robustness to external perturbations, provide an appealing platform for high-density, high-speed, and secure spintronic applications~\cite{Baltz2018,Zelezny2018,Jungwirth2018,Smejkal2018}. Magneto-optical techniques offer a contactless, time-resolved probe of spin dynamics~\cite{Mansuripur1995,Zvezdin1997}, and enable ultrafast control of spin configurations via all-optical switching and optically driven spin currents~\cite{Nemec2018}. However, in conventional collinear antiferromagnets, linear magneto-optical responses—such as the Kerr and Faraday effects—are typically suppressed due to symmetry constraints arising from combined $\mathcal{PT}$ or $\mathcal{T\tau}$ operations, where $\mathcal{P}$, $\mathcal{T}$, and $\tau$ denote spatial inversion, time reversal, and fractional translation, respectively~\cite{P-Yang2022,P-Yang2023}. Circumventing these constraints requires accessing higher-order optical processes. Nonlinear effects such as second harmonic generation, which scales with the square of the electric field, and the Schäfer-Hubert (SH)~\cite{Schafer1990} and Voigt~\cite{Voigt1908} effects, which are quadratic in magnetization, provide viable alternatives. While second harmonic generation is forbidden in centrosymmetric systems, the SH and Voigt effects remains symmetry-allowed and directly probes antiferromagnetic (AFM) order~\footnote{The magneto-optical Voigt and Schäfer-Hubert effects probe polarization rotation in transmission and reflection geometries, respectively. However, the terminology in the literature is often inconsistent: reflection-based measurements are sometimes referred to as the Voigt effect rather than the Schäfer-Hubert effect. In this work, we adopt the latter definition and focus exclusively on the reflection geometry, i.e., the Schäfer-Hubert effect.}. Nonetheless, their typically weak signals pose a major challenge for experimental observation.

Flat-band systems, characterized by nearly dispersionless energy-momentum relations, host electronic states with vanishing group velocity and strong spatial localization. These features enhance Coulomb interactions, stabilizing a rich landscape of correlated phases, including ferromagnetism~\cite{Tasaki1992,Repellin2020,Han2021}, superconductivity~\cite{Cao2018,Balents2020}, Wigner crystallization~\cite{C-Wu2007,Padhi2018}, and fractional Chern insulators~\cite{Xie2021,S-Wu2021}. Their intrinsic tendency to promote magnetic order and enhance interband transitions makes flat bands a natural setting for exploring interaction-driven magneto-optical phenomena. Specifically, the large joint density of states and enhanced transition matrix elements in flat-band systems are expected to boost both linear and nonlinear magneto-optical responses. Despite this promise, magneto-optical effects in flat-band materials remain largely unexplored. Prior studies have primarily focused on linear responses in idealized lattice models, such as the $\alpha$-$\mathcal{T}_3$ model~\cite{Illes2016,Malcolm2016}, Kane model~\cite{Malcolm2014,Akrap2016}, square lattices~\cite{Habibi2022}, and twisted bilayers~\cite{J-Liu2020,Ochoa2020}. In contrast, second-order magneto-optical effects—such as the SH effect—in realistic flat-band AFMs remain virtually unaddressed.

In this work, we employ first-principles calculations combined with group-theoretical analysis to demonstrate that monolayer AFM RuOCl$_2$ exhibits a record-large second-order magneto-optical SH effect, featuring a polarization rotation angle of $28^\circ$. This giant response originates from resonant interband optical transitions involving nearly flat bands. We further find that both the valence and conduction bands exhibit pronounced directional flatness, leading to strongly anisotropic optical absorption along orthogonal crystallographic axes and giving rise to broadband hyperbolic frequency windows spanning the entire visible spectrum. By analyzing dipole selection rules and band topology, we identify the microscopic origin of the anisotropic transitions. This optical anisotropy results in giant linear dichroism (LD) reaching 50\%, surpassing all previously reported values in two-dimensional (2D) magnetic materials. Notably, the SH effect and LD emerge at distinct photon energies, reflecting a momentum-direction–dependent crossover between flat and dispersive bands. Finally, we show that both the SH effect and LD are significantly enhanced with increasing RuOCl$_2$ film thickness. These findings establish a direct link between flat-band physics and large anisotropic magneto-optical responses, opening new avenues for 2D opto-spintronic device applications.

\begin{figure}[t]
	\centering
	\includegraphics[width=1.0\columnwidth]{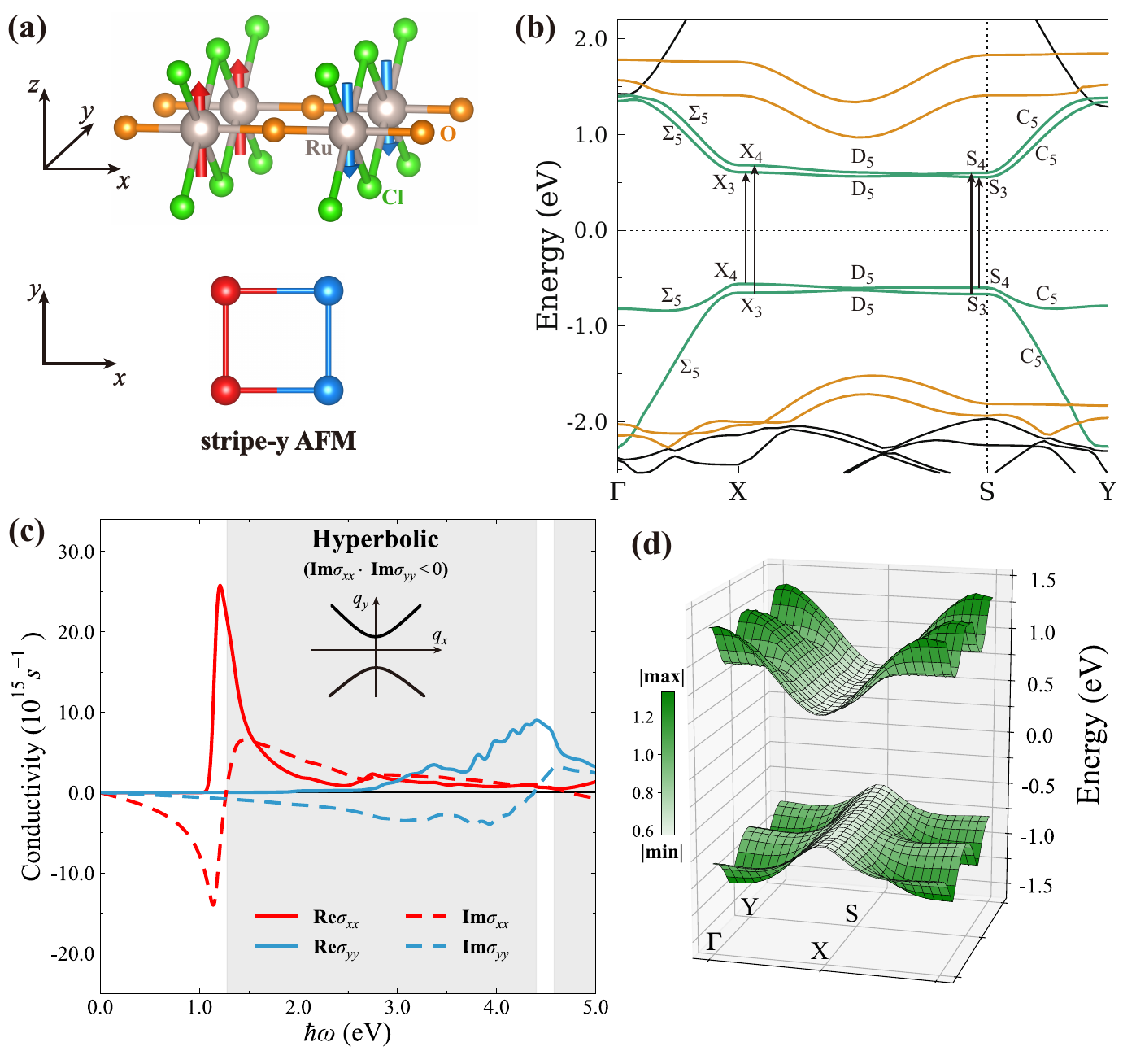}
	\caption{(a) Crystal and magnetic structures of monolayer RuOCl$_2$ with the stripe-$y$ AFM configuration. In the top view, red and blue spheres represent Ru atoms with spin-up and spin-down orientations, respectively; nonmagnetic atoms (O and Cl) are omitted for clarity. (b) Relativistic band structure of monolayer stripe-$y$ AFM RuOCl$_2$, with irreducible co-representations indicated near the band edges. Two groups of bands—exhibiting nearly flat and strongly dispersive behavior along orthogonal crystallographic directions ($\Gamma$–X/S–Y and X–S)—are highlighted in green and gold, respectively. Vertical solid black arrows denote symmetry-allowed optical transitions induced by $x$- and $y$-polarized light at the X and S points. (c) Optical conductivity of monolayer stripe-$y$ AFM RuOCl$_2$. The inset schematically illustrates the hyperbolic dispersion. Shaded regions indicate hyperbolic frequency windows where $\text{Im}\sigma_{xx}$ and $\text{Im}\sigma_{yy}$ have opposite signs. (d) Two-dimensional band structures of the topmost valence and lowest conduction bands. The color map represents the energy difference relative to the band-edge extrema.}
	\label{fig:1}
\end{figure}

Bulk RuOCl$_{2}$ crystallizes in an orthorhombic layered structure with the crystallographic space group $Immm$ (No. 71)~\cite{H-Hillebrecht1997}. In the monolayer form, Ru atoms are connected via a single in-plane O atom along the $x$ direction and via two Cl atoms—positioned above and below the Ru plane—along the $y$ direction [Fig.~\ref{fig:1}(a)]. Several magnetic configurations of monolayer RuOCl$_{2}$ are considered, including the ferromagnetic (FM) and three types of AFM states: stagger, stripe-$x$, and stripe-$y$ (see Fig.~\textcolor{blue}{S1} in the Supplemental Material~\cite{SuppMater}). Total energy calculations reveal that the stripe-$y$ AFM configuration is the most energetically favorable ground state (Supplemental Table~\textcolor{blue}{S1}), which is in good agreement with previous theoretical results~\cite{Y-Zhang2022}. The magnetic space group and the corresponding magnetic point group of the stripe-$y$ AFM state belong to $P_{a}mma$ (No. 51.298) and $mmm.1^\prime$ (No. 8.2.25) respectively. The interlayer binding energy is calculated to be 14.64 meV/\AA$^2$ [Supplemental Fig.~\textcolor{blue}{S2}(a)], which lies within the typical range of 2D layered transition metal dichalcogenides (10--17 meV/\AA$^2$)~\cite{CMO-Bastos2019}. This relatively low binding energy suggests that monolayer RuOCl$_{2}$ can be readily exfoliated from its bulk counterpart. Furthermore, the absence of imaginary frequencies in the phonon spectrum confirms the dynamic stability of the monolayer structure [Supplemental Fig.~\textcolor{blue}{S2}(b)].

After identifying the magnetic ground state, we proceed to investigate the electronic properties. As shown in Fig.~\ref{fig:1}(b), monolayer stripe-$y$ AFM RuOCl$_2$ is a semiconductor with an indirect band gap of 1.11 eV. The valence (conduction) band extremum at the X point is 40 (14) meV higher in energy than that at the S point. Owing to the $\mathcal{PT}$ symmetry within the group $mmm.1^\prime$, all bands are doubly degenerate. The most remarkable feature of the band structure is its pronounced anisotropy. Near the band edges, along the X–S path, the top two valence bands and the bottom two conduction bands are nearly flat, whereas along the $\Gamma$–X and S–Y paths, the bands are highly dispersive. This indicates that charge carriers near the band edges exhibit significantly higher mobility along the Ru–O chains (aligned with the $x$ direction) than along the Ru–Cl chains (aligned with the $y$ direction). Interestingly, at energies further away from the band edges—specifically involving the third and fourth valence and conduction bands—the flat and dispersive characteristics are exchanged between the $x$ and $y$ directions. This implies a corresponding reversal of charge carrier mobility between the two orthogonal crystallographic directions.

\begin{figure*}
	\centering
	\includegraphics[width=2.0\columnwidth]{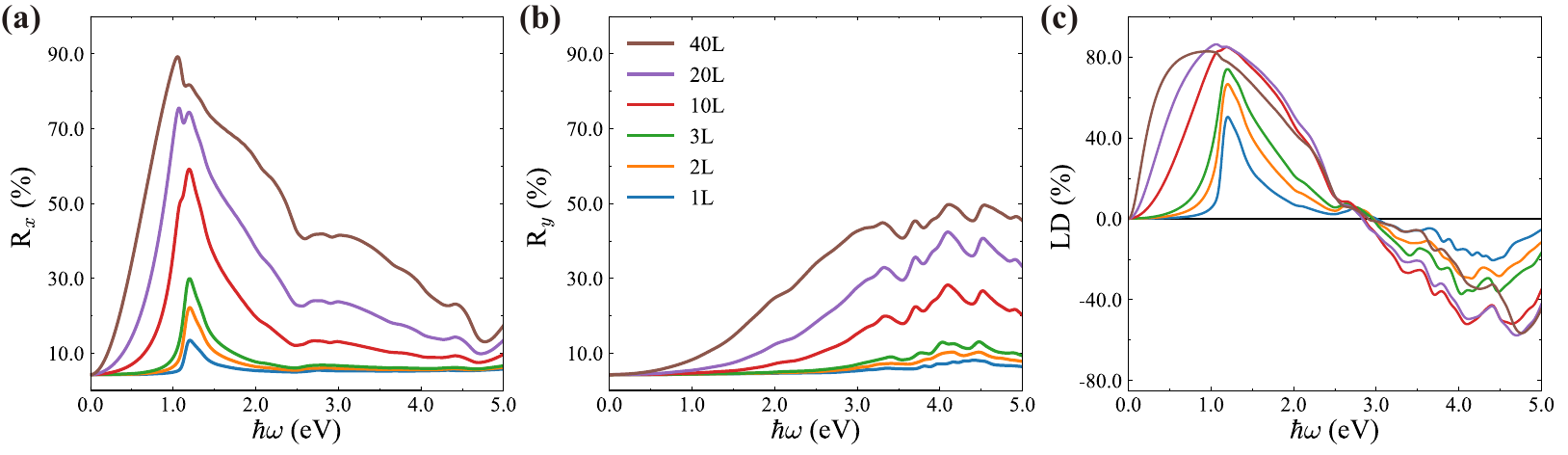}
	\caption{(a,b) Light reflectivity spectra along  orthogonal crystallographic directions ($x$ and $y$) directions and (c) linear dichroism spectra of RuOCl$_2$ thin films with different numbers of layers.}
	\label{fig:2}
\end{figure*}

This strong anisotropy in the electronic structure is reflected in the optical conductivity, which is also highly directional. Due to $\mathcal{PT}$ symmetry, the off-diagonal components of the optical conductivity vanish, as $\mathcal{PT} \sigma_{xy} = -\sigma_{xy}$, leads to $\sigma_{xy} = 0$~\cite{P-Yang2022}. This implies that only the diagonal components need to be considered, as shown in Fig.~\ref{fig:1}(c). The absorptive parts of the optical conductivity—namely, the real parts of $\sigma_{xx}$ and $\sigma_{yy}$—originate from interband transitions. Notably, $\text{Re}\sigma_{xx}$ rises sharply just above the optical gap of 1.11 eV, reaching a pronounced peak at 1.2 eV, while $\text{Re}\sigma_{yy}$ remains negligibly small over the same energy range. According to the Kramers–Kronig relations, the imaginary part of $\sigma_{xx}$ exhibits a deep valley and a broad peak on either side of the $\text{Re}\sigma_{xx}$ peak at 1.2 eV.  At photon energies above approximately 3.0 eV, the anisotropy in conductivity reverses, with both the real and imaginary parts of $\sigma_{yy}$ becoming larger than those of $\sigma_{xx}$. Another remarkable feature of optical conductivity is that the imaginary parts along the $x$ and $y$ directions have opposite signs within the ranges of 1.3–4.4 eV and above 4.6 eV, signaling the emergence of hyperbolic dispersion over a broad spectral window. This hyperbolic window fully covers the visible light range and is broader than that of many other 2D materials (Supplemental Fig.~\textcolor{blue}{S3}), thus indicating monolayer RuOCl$_2$ is a competing 2D hyperbolic material.

The sharp peak in $\text{Re}\sigma_{xx}$ at 1.2 eV is particularly striking and originates from interband transitions between the flat bands near the valence and conduction band edges. Using the MSGCorep package~\cite{G-Liu2021,G-Liu2022}, the calculated irreducible co-representations at the X (S) point are X$_3$ (S$_3$) and X$_4$ (S$_4$), respectively, as labeled in Fig.~\ref{fig:1}(b). According to the dipole selection rules (see Fig.~\textcolor{blue}{S4} and relevant discussion in Supplemental Material), the transitions are symmetry-allowed for incident light polarized along both the $x$ and $y$ direction. Along the high-symmetry paths $\Gamma$–X, X–S, and S–Y, only one irreducible co-representation exists—$\Sigma_5$, D$_5$, and C$_5$, respectively—indicating that all interband transitions along these directions are symmetry-allowed. The $\Gamma$–X and S–Y bands are highly dispersive, implying low effective mass and high carrier mobility along the $x$ direction, which together with fully allowed transitions yields a pronounced optical absorption peak. In contrast, the ultra-flat bands along X–S indicate high effective mass and low mobility along the $y$ direction, resulting in weak absorption despite allowed transitions. As shown in the 2D band structures of the top valence and bottom conduction bands [Fig.~\ref{fig:1}(d)], the bands remain flat not only along X–S but across a broad surrounding region, leading to persistently low optical absorption along the $y$ direction from 1.0 to 3.0 eV. Thus, the strong absorption anisotropy ($\text{Re}\sigma_{xx} \gg \text{Re}\sigma_{yy}$) stems from the contrast between dispersive and flat bands along orthogonal directions.

Notably, the absorption anisotropy reverses at higher photon energies (3.0–5.0 eV), with $\text{Re}\sigma_{xx} \ll \text{Re}\sigma_{yy}$. This stems from a shift in band character: the third and fourth valence and conduction bands become flat along $\Gamma$–X and S–Y, while bands along X–S grow more dispersive. This redistribution of flat and dispersive features across directions drives the reversal in absorption anisotropy.

Building on the anisotropic optical conductivity, we now discuss the resulting giant LD. The reflective LD, quantifying the relative intensity difference between orthogonally polarized lights, is defined as
\begin{equation}\label{LD}
	\text{LD} = \frac{R_x - R_y}{R_x + R_y},
\end{equation}
where $R_{x(y)} = |r_{x(y)}|^2$ is the reflectivity, and $r_{x(y)}$ is the complex field reflectivity at the sample surface, given by $r_{x(y)} = (1 - \tilde{n}_{x(y)}) / (1 + \tilde{n}_{x(y)})$, with $\tilde{n}_{x(y)}$ being the effective refractive index of the thin film on an optically isotropic, nonmagnetic substrate~\cite{P-Yang2023}:
\begin{equation}\label{2D-refractive-1}
	\tilde{n}_{x(y)} = \frac{1 - r_{x(y)}^{\prime} \beta_{x(y)}}{1 + r_{x(y)}^{\prime} \beta_{x(y)}} n_{x(y)}.
\end{equation}
Here, $\beta_{x(y)} = \exp\left(2 i \omega d_\text{eff} n_{x(y)} / c\right)$, with $\omega$ being the light frequency, $c$ the speed of light in vacuum, and $d_\text{eff}$ the effective thickness of the thin film. The interface reflectivity is $r_{x(y)}^{\prime} = (n_{x(y)} - n_s)/(n_{x(y)} + n_s)$, with $n_{x(y)}$ and $n_s$ the refractive indices of the film and substrate, respectively. The film’s refractive index relates to its optical conductivity via $n^2_{x(y)} = 1 + \frac{4\pi i}{\omega} \sigma_{xx(yy)}$, and $n_s$ is obtained from optical constants databases~\cite{refractiveindex}.

\begin{figure*}[t]
	\centering
	\includegraphics[width=2.0\columnwidth]{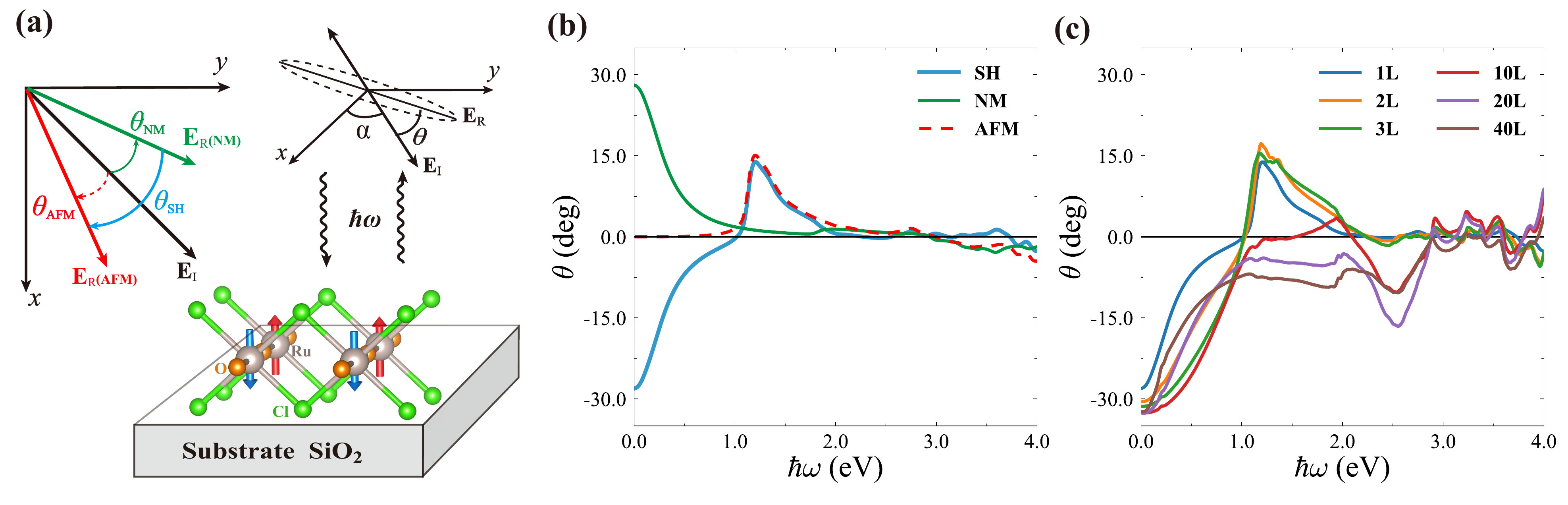}
	\caption{(a) Schematic illustration of the elliptically polarized light reflected from monolayer antiferromagnetic RuOCl$_2$ on a SiO$_2$ substrate. $\textbf{E}_\text{I}$ and $\textbf{E}_\text{R}$ denote the electric fields of the incident and reflected light, respectively. $\alpha$ is the angle between $\textbf{E}_\text{I}$ and the $x$-axis, while $\theta$ represents the rotation angle of the polarization plane of the reflected light relative to the incident light. Specifically, $\textbf{E}_{\text{R(NM)}}$ and $\textbf{E}_{\text{R(AFM)}}$ correspond to the electric fields of the reflected light from NM and AFM RuOCl$_2$, respectively. The associated rotation angles are $\theta_\text{NM}$ and $\theta_\text{AFM}$, and the magneto-optical SH rotation angle is defined as $\theta_\text{SH} = \theta_\text{AFM} - \theta_\text{NM}$. (b) Rotation angle of monolayer NM and AFM RuOCl$_2$, along with the corresponding magneto-optical SH response.  (c) SH rotation angle of RuOCl$_2$ thin films with different numbers of layers.}
	\label{fig:3}
\end{figure*}

The calculated reflectivity and LD spectra for RuOCl$_2$ thin films (from monolayer to 40 layers) on a SiO$_2$ substrate are shown in Fig.~\ref{fig:2}. For the monolayer case, the reflectivity for $x$-polarized light ($R_x$) exhibits a sharp peak of 14\% at 1.2 eV, whereas the reflectivity for $y$-polarized light ($R_y$) is nearly zero, remaining below 5\% in the energy range of 3.0–5.0 eV. This behavior can be understood from the effective refractive index [Eq.~\eqref{2D-refractive-1}], which, for sufficiently thin films, can be approximated as
\begin{equation}\label{2D-refractive-2}
	\tilde{n}_{x(y)} = n_s - \frac{i \omega d_\text{eff}}{c}\left(n_{x(y)}^2 - n_s^2\right).
\end{equation}
Around 1.2 eV, due to the pronounced anisotropy in optical conductivity ($\sigma_{xx} \gg \sigma_{yy}$) [Fig.~\ref{fig:1}(c)], a strongly anisotropic refractive index arises, namely $n_x \gg n_y$. Given the negligibly small $n_y$, the corresponding effective refractive index $\tilde{n}_y$ approaches that of the substrate, $\tilde{n}_y \approx n_s$, as the second term in Eq.~\eqref{2D-refractive-2} becomes negligible at low photon energies and in the monolayer limit. In contrast, the large $n_x$ leads to a significantly larger $\tilde{n}_x$, despite the small values of $\omega$ and $d_\text{eff}$. Consequently, the effective refractive index is also strongly anisotropic, $\tilde{n}_x \gg \tilde{n}_y$, resulting in a pronounced peak in $R_x$ and a negligible $R_y$ around 1.2 eV. This large difference between $R_x$ and $R_y$ produces a remarkable LD of 50.4\% at 1.2 eV, as shown in Fig.~\ref{fig:2}(c). As the photon energy increases above 3.0 eV, the situation reverses, with $R_x < R_y$, which can be attributed to the reversal of optical conductivity anisotropy ($\sigma_{xx} \ll \sigma_{yy}$) [Fig.~\ref{fig:1}(c)]. Particularly, at 4.4 eV, the LD reaches as much as –20.1\%. Benefiting from this energy-dependent strong LD, the crystal orientation of a monolayer can be readily identified using azimuth-dependent reflectance difference microscopy~\cite{J-Wang2022}.

The reflectivity is monotonically enhanced by increasing the film thickness, while the LD presents a slightly complicated dependence on the film thickness. As the thickness increases from monolayer to 10 layers, $R_x$ grows much faster than $R_y$, particularly in the low-energy range, below 3.0 eV [Figs.~\ref{fig:2}(a) and~\ref{fig:2}(b)]. As a result, the LD not only reaches a maximum of 84.7\% at 1.2 eV but is also significantly enhanced across the visible light (e.g., 1.6–2.5 eV) and near-infrared (0.4--0.9 eV) regions [Fig.~\ref{fig:2}(c)]. With further increases in thickness from 20 to 40 layers, the peak value of LD remains unchanged, but the LD stays relatively large across a broad energy range, particularly extending into the far-infrared region below 0.4 eV. The thickness dependence of LD is also evident in the high-energy range, specifically between 3.0 and 5.0 eV, where the LD reaches up to -57.6\% at 4.7 eV [Fig.~\ref{fig:2}(c)].  The LD observed in RuOCl$_2$ reaches a record-high value, surpassing that of all other known 2D materials (Supplemental Table~\textcolor{blue}{S2}).

Along with the strong LD, the reflected light can become elliptically polarized, accompanied by a rotation of the polarization plane relative to the incident light, as depicted in Fig.~\ref{fig:3}(a). When the electric field of the incident light ($\textbf{E}_\text{I}$) is oriented at an angle $\alpha = 45^{\circ}$ with respect to the $x$-axis, both the rotation angle ($\theta$) and the ellipticity ($\psi$) reach their maximum values, given by~\cite{P-Yang2023}:
\begin{align}
		\theta &= \frac{\pi}{4} - \frac{1}{2} \operatorname{atan}\left(\frac{2 \operatorname{Re} \chi}{1-|\chi|^{2}}\right) \frac{\pi}{4}, \\
		\psi &= \frac{1}{2} \operatorname{asin}\left(\frac{2 \operatorname{Im} \chi}{1+|\chi|^{2}}\right), 
\end{align}
where $\chi = r_{y}/r_{x}$. A positive (negative) $\theta$ corresponds to a clockwise (counterclockwise) rotation, and a positive (negative) $\psi$ indicates left- (right-) handed elliptical polarization. The rotation of the polarization plane and the corresponding ellipticity induced by magnetic order are referred to as the magneto-optical SH effect, whereas the contributions arising from crystal anisotropy are termed natural linear birefringence (NLB). In stripe-$y$ AFM RuOCl$_2$, due to unequal in-plane lattice constants (Supplemental Table~\textcolor{blue}{S1}), both the SH effect and NLB coexist. The NLB contribution can be evaluated in RuOCl$_2$ by artificially removing its magnetic order, i.e., by considering the system in a nonmagnetic (NM) state. Subsequently, the SH rotation angle is obtained by subtracting the NLB contribution from the results of AFM state~\cite{P-Yang2022}:
\begin{equation}\label{angle}
		\theta_\text{SH} = \theta_\text{AFM} - \theta_\text{NM}.
\end{equation}

The SH rotation angle of a monolayer RuOCl$_2$ on a SiO$_2$ substrate is shown in Fig.~\ref{fig:3}(b). In the AFM case, the rotation angle exhibits a frequency dependence similar to that of the LD spectrum, reaching a maximum of 15.1$^\circ$ at 1.2 eV. This behavior essentially originates from the anisotropy in the optical conductivity. As derived in our previous work~\cite{P-Yang2023}, the rotation of polarization plane is related to the difference in optical conductivity between the $x$ and $y$ directions, following the relation $\theta + i\psi \propto |\sigma_{xx} - \sigma_{yy}|$. The large difference in the real parts of $\sigma_{xx}$ and $\sigma_{yy}$ near 1.2 eV gives rise to the pronounced peak in $\theta_\text{AFM}$.  In the NM case, the rotation angle $\theta_\text{NM}$ exhibits a sharp peak of 28.0$^\circ$ close to the zero-frequency limit. This is attributed to the metallic nature of monolayer NM RuOCl$_2$, which shows strong anisotropy in the intraband contributions to the reflectivity and LD below 0.5 eV (Supplemental Fig.~\textcolor{blue}{S5}). Given that the SH angle is defined as the difference between the rotation angles in AFM and NM states [Eq.~\eqref{angle}], and that $\theta_\text{AFM}$ is nearly zero below 1.0 eV, the SH angle in this low-energy regime is dominated by and opposite in sign to $\theta_\text{NM}$.  Conversely, for photon energies above 1.0 eV, the optical response in the NM state becomes nearly isotropic, and the SH angle closely follows $\theta_\text{AFM}$. Notably, the SH angle in RuOCl$_2$ is several orders of magnitude larger than those observed in typical 2D and bulk magnetic materials (Supplemental Table~\textcolor{blue}{S2}). Figure~\ref{fig:3}(c) shows the dependence of SH rotation angles on the thickness of RuOCl$_2$ thin films. For films thinner than three layers, the SH angle profile resembles that of the monolayer, with the exception of a gradual enhancement in the 0–1.0 eV energy range. Starting from ten layers, the peak around 1.2 eV vanishes, and a new peak emerges, centered at 2.5 eV.

In summary, we have theoretically uncovered a giant second-order magneto-optical SH effect in monolayer AFM RuOCl$_2$, driven by flat-band–enhanced interband optical transitions. Both the valence and conduction bands exhibit pronounced directional flatness, leading to strongly anisotropic optical absorption along orthogonal crystallographic axes and broadband hyperbolic frequency windows spanning the entire visible range. The material also hosts record-high LD, reaching 55\%, surpassing all previously reported values in 2D systems. The SH and LD effects emerge at distinct photon energies, reflecting a momentum-direction–dependent crossover between flat and dispersive bands. Both effects persist and are further enhanced in few-layer films. Importantly, the underlying anisotropic flat-band electronic structure is not unique to RuOCl$_2$; similar features have been identified in other 2D magnets, including Eu$_2$SiO$_7$Cl, Fe$_3$Sb$_2$O$_4$Br$_4$, and Mn$_3$Sb$_2$O$_4$I$_4$~\cite{J-Duan2024}. Our findings establish 2D flat-band magnets as a rare class of materials where magnetic order and flat-band physics coexist to produce highly tunable, directionally selective magneto-optical responses—opening new avenues for next-generation optical modulators and integrated opto-spintronic technologies.

The authors thank Yinong Zhou (UCI), Jun Ding (HUE), Chengwang Niu (SDU), and Jinzhong Zhang (ECNU) for helpful discussion. This work is supported by the National Natural Science Foundation of China (Grants No. 12274027, No. 12234003, and No. 12321004), the National Key R\&D Program of China (Grants No. 2022YFA1402600, No. 2022YFA1403800, and No. 2020YFA0308800), and the Fundamental Research Funds for the Central Universities (Grant No. 2024CX06104).

\bibliography{references}

\begin{thebibliography}{37}%
\makeatletter
\providecommand \@ifxundefined [1]{%
 \@ifx{#1\undefined}
}%
\providecommand \@ifnum [1]{%
 \ifnum #1\expandafter \@firstoftwo
 \else \expandafter \@secondoftwo
 \fi
}%
\providecommand \@ifx [1]{%
 \ifx #1\expandafter \@firstoftwo
 \else \expandafter \@secondoftwo
 \fi
}%
\providecommand \natexlab [1]{#1}%
\providecommand \enquote  [1]{``#1''}%
\providecommand \bibnamefont  [1]{#1}%
\providecommand \bibfnamefont [1]{#1}%
\providecommand \citenamefont [1]{#1}%
\providecommand \href@noop [0]{\@secondoftwo}%
\providecommand \href [0]{\begingroup \@sanitize@url \@href}%
\providecommand \@href[1]{\@@startlink{#1}\@@href}%
\providecommand \@@href[1]{\endgroup#1\@@endlink}%
\providecommand \@sanitize@url [0]{\catcode `\\12\catcode `\$12\catcode
  `\&12\catcode `\#12\catcode `\^12\catcode `\_12\catcode `\%12\relax}%
\providecommand \@@startlink[1]{}%
\providecommand \@@endlink[0]{}%
\providecommand \url  [0]{\begingroup\@sanitize@url \@url }%
\providecommand \@url [1]{\endgroup\@href {#1}{\urlprefix }}%
\providecommand \urlprefix  [0]{URL }%
\providecommand \Eprint [0]{\href }%
\providecommand \doibase [0]{http://dx.doi.org/}%
\providecommand \selectlanguage [0]{\@gobble}%
\providecommand \bibinfo  [0]{\@secondoftwo}%
\providecommand \bibfield  [0]{\@secondoftwo}%
\providecommand \translation [1]{[#1]}%
\providecommand \BibitemOpen [0]{}%
\providecommand \bibitemStop [0]{}%
\providecommand \bibitemNoStop [0]{.\EOS\space}%
\providecommand \EOS [0]{\spacefactor3000\relax}%
\providecommand \BibitemShut  [1]{\csname bibitem#1\endcsname}%
\let\auto@bib@innerbib\@empty
\bibitem [{\citenamefont {Baltz}\ \emph {et~al.}(2018)\citenamefont {Baltz},
  \citenamefont {Manchon}, \citenamefont {Tsoi}, \citenamefont {Moriyama},
  \citenamefont {Ono},\ and\ \citenamefont {Tserkovnyak}}]{Baltz2018}%
  \BibitemOpen
  \bibfield  {author} {\bibinfo {author} {\bibfnamefont {V.}~\bibnamefont
  {Baltz}}, \bibinfo {author} {\bibfnamefont {A.}~\bibnamefont {Manchon}},
  \bibinfo {author} {\bibfnamefont {M.}~\bibnamefont {Tsoi}}, \bibinfo {author}
  {\bibfnamefont {T.}~\bibnamefont {Moriyama}}, \bibinfo {author}
  {\bibfnamefont {T.}~\bibnamefont {Ono}}, \ and\ \bibinfo {author}
  {\bibfnamefont {Y.}~\bibnamefont {Tserkovnyak}},\ }\href
  {https://link.aps.org/doi/10.1103/RevModPhys.90.015005} {\bibfield  {journal}
  {\bibinfo  {journal} {Rev. Mod. Phys.}\ }\textbf {\bibinfo {volume} {90}},\
  \bibinfo {pages} {015005} (\bibinfo {year} {2018})}\BibitemShut {NoStop}%
\bibitem [{\citenamefont {{\v Z}elezn{\'y}}\ \emph {et~al.}(2018)\citenamefont
  {{\v Z}elezn{\'y}}, \citenamefont {Wadley}, \citenamefont {Olejn{\'{\i}}k},
  \citenamefont {Hoffmann},\ and\ \citenamefont {Ohno}}]{Zelezny2018}%
  \BibitemOpen
  \bibfield  {author} {\bibinfo {author} {\bibfnamefont {J.}~\bibnamefont {{\v
  Z}elezn{\'y}}}, \bibinfo {author} {\bibfnamefont {P.}~\bibnamefont {Wadley}},
  \bibinfo {author} {\bibfnamefont {K.}~\bibnamefont {Olejn{\'{\i}}k}},
  \bibinfo {author} {\bibfnamefont {A.}~\bibnamefont {Hoffmann}}, \ and\
  \bibinfo {author} {\bibfnamefont {H.}~\bibnamefont {Ohno}},\ }\href {\doibase
  10.1038/s41567-018-0062-7} {\bibfield  {journal} {\bibinfo  {journal} {Nat.
  Phys.}\ }\textbf {\bibinfo {volume} {14}},\ \bibinfo {pages} {220} (\bibinfo
  {year} {2018})}\BibitemShut {NoStop}%
\bibitem [{\citenamefont {Jungwirth}\ \emph {et~al.}(2018)\citenamefont
  {Jungwirth}, \citenamefont {Sinova}, \citenamefont {Manchon}, \citenamefont
  {Marti}, \citenamefont {Wunderlich},\ and\ \citenamefont
  {Felser}}]{Jungwirth2018}%
  \BibitemOpen
  \bibfield  {author} {\bibinfo {author} {\bibfnamefont {T.}~\bibnamefont
  {Jungwirth}}, \bibinfo {author} {\bibfnamefont {J.}~\bibnamefont {Sinova}},
  \bibinfo {author} {\bibfnamefont {A.}~\bibnamefont {Manchon}}, \bibinfo
  {author} {\bibfnamefont {X.}~\bibnamefont {Marti}}, \bibinfo {author}
  {\bibfnamefont {J.}~\bibnamefont {Wunderlich}}, \ and\ \bibinfo {author}
  {\bibfnamefont {C.}~\bibnamefont {Felser}},\ }\href {\doibase
  10.1038/s41567-018-0063-6} {\bibfield  {journal} {\bibinfo  {journal} {Nat.
  Phys.}\ }\textbf {\bibinfo {volume} {14}},\ \bibinfo {pages} {200} (\bibinfo
  {year} {2018})}\BibitemShut {NoStop}%
\bibitem [{\citenamefont {{\v S}mejkal}\ \emph {et~al.}(2018)\citenamefont {{\v
  S}mejkal}, \citenamefont {Mokrousov}, \citenamefont {Yan},\ and\
  \citenamefont {MacDonald}}]{Smejkal2018}%
  \BibitemOpen
  \bibfield  {author} {\bibinfo {author} {\bibfnamefont {L.}~\bibnamefont {{\v
  S}mejkal}}, \bibinfo {author} {\bibfnamefont {Y.}~\bibnamefont {Mokrousov}},
  \bibinfo {author} {\bibfnamefont {B.}~\bibnamefont {Yan}}, \ and\ \bibinfo
  {author} {\bibfnamefont {A.~H.}\ \bibnamefont {MacDonald}},\ }\href {\doibase
  10.1038/s41567-018-0064-5} {\bibfield  {journal} {\bibinfo  {journal} {Nat.
  Phys.}\ }\textbf {\bibinfo {volume} {14}},\ \bibinfo {pages} {242} (\bibinfo
  {year} {2018})}\BibitemShut {NoStop}%
\bibitem [{\citenamefont {Mansuripur}(1995)}]{Mansuripur1995}%
  \BibitemOpen
  \bibfield  {author} {\bibinfo {author} {\bibfnamefont {M.}~\bibnamefont
  {Mansuripur}},\ }\href@noop {} {\emph {\bibinfo {title} {The Physical
  Principles of Magneto-Optical Recording}}}\ (\bibinfo  {publisher} {Cambridge
  University Press, New York},\ \bibinfo {year} {1995})\BibitemShut {NoStop}%
\bibitem [{\citenamefont {Zvezdin}\ and\ \citenamefont
  {Kotov}(1997)}]{Zvezdin1997}%
  \BibitemOpen
  \bibfield  {author} {\bibinfo {author} {\bibfnamefont {A.~K.}\ \bibnamefont
  {Zvezdin}}\ and\ \bibinfo {author} {\bibfnamefont {V.~A.}\ \bibnamefont
  {Kotov}},\ }\href@noop {} {\emph {\bibinfo {title} {Modern Magnetooptics and
  Magnetooptical Materials}}}\ (\bibinfo  {publisher} {Institute of Physics
  Publishing, Bristol and Philadelphia},\ \bibinfo {year} {1997})\BibitemShut
  {NoStop}%
\bibitem [{\citenamefont {N{\v e}mec}\ \emph {et~al.}(2018)\citenamefont {N{\v
  e}mec}, \citenamefont {Fiebig}, \citenamefont {Kampfrath},\ and\
  \citenamefont {Kimel}}]{Nemec2018}%
  \BibitemOpen
  \bibfield  {author} {\bibinfo {author} {\bibfnamefont {P.}~\bibnamefont {N{\v
  e}mec}}, \bibinfo {author} {\bibfnamefont {M.}~\bibnamefont {Fiebig}},
  \bibinfo {author} {\bibfnamefont {T.}~\bibnamefont {Kampfrath}}, \ and\
  \bibinfo {author} {\bibfnamefont {A.~V.}\ \bibnamefont {Kimel}},\ }\href
  {\doibase 10.1038/s41567-018-0051-x} {\bibfield  {journal} {\bibinfo
  {journal} {Nat. Phys.}\ }\textbf {\bibinfo {volume} {14}},\ \bibinfo {pages}
  {229} (\bibinfo {year} {2018})}\BibitemShut {NoStop}%
\bibitem [{\citenamefont {Yang}\ \emph {et~al.}(2022)\citenamefont {Yang},
  \citenamefont {Feng}, \citenamefont {Zhou}, \citenamefont {Yang},\ and\
  \citenamefont {Yao}}]{P-Yang2022}%
  \BibitemOpen
  \bibfield  {author} {\bibinfo {author} {\bibfnamefont {P.}~\bibnamefont
  {Yang}}, \bibinfo {author} {\bibfnamefont {W.}~\bibnamefont {Feng}}, \bibinfo
  {author} {\bibfnamefont {X.}~\bibnamefont {Zhou}}, \bibinfo {author}
  {\bibfnamefont {X.}~\bibnamefont {Yang}}, \ and\ \bibinfo {author}
  {\bibfnamefont {Y.}~\bibnamefont {Yao}},\ }\href {\doibase
  10.1103/PhysRevB.106.174427} {\bibfield  {journal} {\bibinfo  {journal}
  {Phys. Rev. B}\ }\textbf {\bibinfo {volume} {106}},\ \bibinfo {pages}
  {174427} (\bibinfo {year} {2022})}\BibitemShut {NoStop}%
\bibitem [{\citenamefont {Yang}\ \emph {et~al.}(2023)\citenamefont {Yang},
  \citenamefont {Feng}, \citenamefont {Liu}, \citenamefont {Guo},\ and\
  \citenamefont {Yao}}]{P-Yang2023}%
  \BibitemOpen
  \bibfield  {author} {\bibinfo {author} {\bibfnamefont {P.}~\bibnamefont
  {Yang}}, \bibinfo {author} {\bibfnamefont {W.}~\bibnamefont {Feng}}, \bibinfo
  {author} {\bibfnamefont {G.-B.}\ \bibnamefont {Liu}}, \bibinfo {author}
  {\bibfnamefont {G.-Y.}\ \bibnamefont {Guo}}, \ and\ \bibinfo {author}
  {\bibfnamefont {Y.}~\bibnamefont {Yao}},\ }\href {\doibase
  10.1103/PhysRevB.107.214437} {\bibfield  {journal} {\bibinfo  {journal}
  {Phys. Rev. B}\ }\textbf {\bibinfo {volume} {107}},\ \bibinfo {pages}
  {214437} (\bibinfo {year} {2023})}\BibitemShut {NoStop}%
\bibitem [{\citenamefont {Sch{\"a}fer}\ and\ \citenamefont
  {Hubert}(1990)}]{Schafer1990}%
  \BibitemOpen
  \bibfield  {author} {\bibinfo {author} {\bibfnamefont {R.}~\bibnamefont
  {Sch{\"a}fer}}\ and\ \bibinfo {author} {\bibfnamefont {A.}~\bibnamefont
  {Hubert}},\ }\href {\doibase https://doi.org/10.1002/pssa.2211180131}
  {\bibfield  {journal} {\bibinfo  {journal} {Phys. Stat. Sol. (a)}\ }\textbf
  {\bibinfo {volume} {118}},\ \bibinfo {pages} {271} (\bibinfo {year}
  {1990})}\BibitemShut {NoStop}%
\bibitem [{\citenamefont {Voigt}(1908)}]{Voigt1908}%
  \BibitemOpen
  \bibfield  {author} {\bibinfo {author} {\bibfnamefont {W.}~\bibnamefont
  {Voigt}},\ }\href@noop {} {\emph {\bibinfo {title} {Magneto-und
  Elektrooptik}}}\ (\bibinfo  {publisher} {Leipzig, B. G. Teubner},\ \bibinfo
  {year} {1908})\BibitemShut {NoStop}%
\bibitem [{Note1()}]{Note1}%
  \BibitemOpen
  \bibinfo {note} {The magneto-optical Voigt and Schäfer-Hubert effects probe
  polarization rotation in transmission and reflection geometries,
  respectively. However, the terminology in the literature is often
  inconsistent: reflection-based measurements are sometimes referred to as the
  Voigt effect rather than the Schäfer-Hubert effect. In this work, we adopt
  the latter definition and focus exclusively on the reflection geometry, i.e.,
  the Schäfer-Hubert effect.}\BibitemShut {Stop}%
\bibitem [{\citenamefont {Tasaki}(1992)}]{Tasaki1992}%
  \BibitemOpen
  \bibfield  {author} {\bibinfo {author} {\bibfnamefont {H.}~\bibnamefont
  {Tasaki}},\ }\href {\doibase 10.1103/PhysRevLett.69.1608} {\bibfield
  {journal} {\bibinfo  {journal} {Phys. Rev. Lett.}\ }\textbf {\bibinfo
  {volume} {69}},\ \bibinfo {pages} {1608} (\bibinfo {year}
  {1992})}\BibitemShut {NoStop}%
\bibitem [{\citenamefont {Repellin}\ \emph {et~al.}(2020)\citenamefont
  {Repellin}, \citenamefont {Dong}, \citenamefont {Zhang},\ and\ \citenamefont
  {Senthil}}]{Repellin2020}%
  \BibitemOpen
  \bibfield  {author} {\bibinfo {author} {\bibfnamefont {C.}~\bibnamefont
  {Repellin}}, \bibinfo {author} {\bibfnamefont {Z.}~\bibnamefont {Dong}},
  \bibinfo {author} {\bibfnamefont {Y.-H.}\ \bibnamefont {Zhang}}, \ and\
  \bibinfo {author} {\bibfnamefont {T.}~\bibnamefont {Senthil}},\ }\href
  {\doibase 10.1103/PhysRevLett.124.187601} {\bibfield  {journal} {\bibinfo
  {journal} {Phys. Rev. Lett.}\ }\textbf {\bibinfo {volume} {124}},\ \bibinfo
  {pages} {187601} (\bibinfo {year} {2020})}\BibitemShut {NoStop}%
\bibitem [{\citenamefont {Han}\ \emph {et~al.}(2021)\citenamefont {Han},
  \citenamefont {Inoue}, \citenamefont {Fang}, \citenamefont {John},
  \citenamefont {Ye}, \citenamefont {Chan}, \citenamefont {Graf}, \citenamefont
  {Suzuki}, \citenamefont {Ghimire}, \citenamefont {Cho}, \citenamefont
  {Kaxiras},\ and\ \citenamefont {Checkelsky}}]{Han2021}%
  \BibitemOpen
  \bibfield  {author} {\bibinfo {author} {\bibfnamefont {M.}~\bibnamefont
  {Han}}, \bibinfo {author} {\bibfnamefont {H.}~\bibnamefont {Inoue}}, \bibinfo
  {author} {\bibfnamefont {S.}~\bibnamefont {Fang}}, \bibinfo {author}
  {\bibfnamefont {C.}~\bibnamefont {John}}, \bibinfo {author} {\bibfnamefont
  {L.}~\bibnamefont {Ye}}, \bibinfo {author} {\bibfnamefont {M.~K.}\
  \bibnamefont {Chan}}, \bibinfo {author} {\bibfnamefont {D.}~\bibnamefont
  {Graf}}, \bibinfo {author} {\bibfnamefont {T.}~\bibnamefont {Suzuki}},
  \bibinfo {author} {\bibfnamefont {M.~P.}\ \bibnamefont {Ghimire}}, \bibinfo
  {author} {\bibfnamefont {W.~J.}\ \bibnamefont {Cho}}, \bibinfo {author}
  {\bibfnamefont {E.}~\bibnamefont {Kaxiras}}, \ and\ \bibinfo {author}
  {\bibfnamefont {J.~G.}\ \bibnamefont {Checkelsky}},\ }\href {\doibase
  10.1038/s41467-021-25705-1} {\bibfield  {journal} {\bibinfo  {journal} {Nat.
  Commun.}\ }\textbf {\bibinfo {volume} {12}},\ \bibinfo {pages} {5345}
  (\bibinfo {year} {2021})}\BibitemShut {NoStop}%
\bibitem [{\citenamefont {Cao}\ \emph {et~al.}(2018)\citenamefont {Cao},
  \citenamefont {Fatemi}, \citenamefont {Fang}, \citenamefont {Watanabe},
  \citenamefont {Taniguchi}, \citenamefont {Kaxiras},\ and\ \citenamefont
  {Jarillo-Herrero}}]{Cao2018}%
  \BibitemOpen
  \bibfield  {author} {\bibinfo {author} {\bibfnamefont {Y.}~\bibnamefont
  {Cao}}, \bibinfo {author} {\bibfnamefont {V.}~\bibnamefont {Fatemi}},
  \bibinfo {author} {\bibfnamefont {S.}~\bibnamefont {Fang}}, \bibinfo {author}
  {\bibfnamefont {K.}~\bibnamefont {Watanabe}}, \bibinfo {author}
  {\bibfnamefont {T.}~\bibnamefont {Taniguchi}}, \bibinfo {author}
  {\bibfnamefont {E.}~\bibnamefont {Kaxiras}}, \ and\ \bibinfo {author}
  {\bibfnamefont {P.}~\bibnamefont {Jarillo-Herrero}},\ }\href {\doibase
  10.1038/nature26160} {\bibfield  {journal} {\bibinfo  {journal} {Nature}\
  }\textbf {\bibinfo {volume} {556}},\ \bibinfo {pages} {43} (\bibinfo {year}
  {2018})}\BibitemShut {NoStop}%
\bibitem [{\citenamefont {Balents}\ \emph {et~al.}(2020)\citenamefont
  {Balents}, \citenamefont {Dean}, \citenamefont {Efetov},\ and\ \citenamefont
  {Young}}]{Balents2020}%
  \BibitemOpen
  \bibfield  {author} {\bibinfo {author} {\bibfnamefont {L.}~\bibnamefont
  {Balents}}, \bibinfo {author} {\bibfnamefont {C.~R.}\ \bibnamefont {Dean}},
  \bibinfo {author} {\bibfnamefont {D.~K.}\ \bibnamefont {Efetov}}, \ and\
  \bibinfo {author} {\bibfnamefont {A.~F.}\ \bibnamefont {Young}},\ }\href
  {\doibase 10.1038/s41567-020-0906-9} {\bibfield  {journal} {\bibinfo
  {journal} {Nat. Phys.}\ }\textbf {\bibinfo {volume} {16}},\ \bibinfo {pages}
  {725} (\bibinfo {year} {2020})}\BibitemShut {NoStop}%
\bibitem [{\citenamefont {Wu}\ \emph {et~al.}(2007)\citenamefont {Wu},
  \citenamefont {Bergman}, \citenamefont {Balents},\ and\ \citenamefont
  {Das~Sarma}}]{C-Wu2007}%
  \BibitemOpen
  \bibfield  {author} {\bibinfo {author} {\bibfnamefont {C.}~\bibnamefont
  {Wu}}, \bibinfo {author} {\bibfnamefont {D.}~\bibnamefont {Bergman}},
  \bibinfo {author} {\bibfnamefont {L.}~\bibnamefont {Balents}}, \ and\
  \bibinfo {author} {\bibfnamefont {S.}~\bibnamefont {Das~Sarma}},\ }\href
  {\doibase 10.1103/PhysRevLett.99.070401} {\bibfield  {journal} {\bibinfo
  {journal} {Phys. Rev. Lett.}\ }\textbf {\bibinfo {volume} {99}},\ \bibinfo
  {pages} {070401} (\bibinfo {year} {2007})}\BibitemShut {NoStop}%
\bibitem [{\citenamefont {Padhi}\ \emph {et~al.}(2018)\citenamefont {Padhi},
  \citenamefont {Setty},\ and\ \citenamefont {Phillips}}]{Padhi2018}%
  \BibitemOpen
  \bibfield  {author} {\bibinfo {author} {\bibfnamefont {B.}~\bibnamefont
  {Padhi}}, \bibinfo {author} {\bibfnamefont {C.}~\bibnamefont {Setty}}, \ and\
  \bibinfo {author} {\bibfnamefont {P.~W.}\ \bibnamefont {Phillips}},\ }\href
  {\doibase 10.1021/acs.nanolett.8b02033} {\bibfield  {journal} {\bibinfo
  {journal} {Nano Lett.}\ }\textbf {\bibinfo {volume} {18}},\ \bibinfo {pages}
  {6175} (\bibinfo {year} {2018})}\BibitemShut {NoStop}%
\bibitem [{\citenamefont {Xie}\ \emph {et~al.}(2021)\citenamefont {Xie},
  \citenamefont {Pierce}, \citenamefont {Park}, \citenamefont {Parker},
  \citenamefont {Khalaf}, \citenamefont {Ledwith}, \citenamefont {Cao},
  \citenamefont {Lee}, \citenamefont {Chen}, \citenamefont {Forrester},
  \citenamefont {Watanabe}, \citenamefont {Taniguchi}, \citenamefont
  {Vishwanath}, \citenamefont {Jarillo-Herrero},\ and\ \citenamefont
  {Yacoby}}]{Xie2021}%
  \BibitemOpen
  \bibfield  {author} {\bibinfo {author} {\bibfnamefont {Y.}~\bibnamefont
  {Xie}}, \bibinfo {author} {\bibfnamefont {A.~T.}\ \bibnamefont {Pierce}},
  \bibinfo {author} {\bibfnamefont {J.~M.}\ \bibnamefont {Park}}, \bibinfo
  {author} {\bibfnamefont {D.~E.}\ \bibnamefont {Parker}}, \bibinfo {author}
  {\bibfnamefont {E.}~\bibnamefont {Khalaf}}, \bibinfo {author} {\bibfnamefont
  {P.}~\bibnamefont {Ledwith}}, \bibinfo {author} {\bibfnamefont
  {Y.}~\bibnamefont {Cao}}, \bibinfo {author} {\bibfnamefont {S.~H.}\
  \bibnamefont {Lee}}, \bibinfo {author} {\bibfnamefont {S.}~\bibnamefont
  {Chen}}, \bibinfo {author} {\bibfnamefont {P.~R.}\ \bibnamefont {Forrester}},
  \bibinfo {author} {\bibfnamefont {K.}~\bibnamefont {Watanabe}}, \bibinfo
  {author} {\bibfnamefont {T.}~\bibnamefont {Taniguchi}}, \bibinfo {author}
  {\bibfnamefont {A.}~\bibnamefont {Vishwanath}}, \bibinfo {author}
  {\bibfnamefont {P.}~\bibnamefont {Jarillo-Herrero}}, \ and\ \bibinfo {author}
  {\bibfnamefont {A.}~\bibnamefont {Yacoby}},\ }\href {\doibase
  10.1038/s41586-021-04002-3} {\bibfield  {journal} {\bibinfo  {journal}
  {Nature}\ }\textbf {\bibinfo {volume} {600}},\ \bibinfo {pages} {439}
  (\bibinfo {year} {2021})}\BibitemShut {NoStop}%
\bibitem [{\citenamefont {Wu}\ \emph {et~al.}(2021)\citenamefont {Wu},
  \citenamefont {Zhang}, \citenamefont {Watanabe}, \citenamefont {Taniguchi},\
  and\ \citenamefont {Andrei}}]{S-Wu2021}%
  \BibitemOpen
  \bibfield  {author} {\bibinfo {author} {\bibfnamefont {S.}~\bibnamefont
  {Wu}}, \bibinfo {author} {\bibfnamefont {Z.}~\bibnamefont {Zhang}}, \bibinfo
  {author} {\bibfnamefont {K.}~\bibnamefont {Watanabe}}, \bibinfo {author}
  {\bibfnamefont {T.}~\bibnamefont {Taniguchi}}, \ and\ \bibinfo {author}
  {\bibfnamefont {E.~Y.}\ \bibnamefont {Andrei}},\ }\href {\doibase
  10.1038/s41563-020-00911-2} {\bibfield  {journal} {\bibinfo  {journal} {Nat.
  Mater.}\ }\textbf {\bibinfo {volume} {20}},\ \bibinfo {pages} {488} (\bibinfo
  {year} {2021})}\BibitemShut {NoStop}%
\bibitem [{\citenamefont {Illes}\ and\ \citenamefont
  {Nicol}(2016)}]{Illes2016}%
  \BibitemOpen
  \bibfield  {author} {\bibinfo {author} {\bibfnamefont {E.}~\bibnamefont
  {Illes}}\ and\ \bibinfo {author} {\bibfnamefont {E.~J.}\ \bibnamefont
  {Nicol}},\ }\href {\doibase 10.1103/PhysRevB.94.125435} {\bibfield  {journal}
  {\bibinfo  {journal} {Phys. Rev. B}\ }\textbf {\bibinfo {volume} {94}},\
  \bibinfo {pages} {125435} (\bibinfo {year} {2016})}\BibitemShut {NoStop}%
\bibitem [{\citenamefont {Malcolm}\ and\ \citenamefont
  {Nicol}(2016)}]{Malcolm2016}%
  \BibitemOpen
  \bibfield  {author} {\bibinfo {author} {\bibfnamefont {J.~D.}\ \bibnamefont
  {Malcolm}}\ and\ \bibinfo {author} {\bibfnamefont {E.~J.}\ \bibnamefont
  {Nicol}},\ }\href {\doibase 10.1103/PhysRevB.94.224305} {\bibfield  {journal}
  {\bibinfo  {journal} {Phys. Rev. B}\ }\textbf {\bibinfo {volume} {94}},\
  \bibinfo {pages} {224305} (\bibinfo {year} {2016})}\BibitemShut {NoStop}%
\bibitem [{\citenamefont {Malcolm}\ and\ \citenamefont
  {Nicol}(2014)}]{Malcolm2014}%
  \BibitemOpen
  \bibfield  {author} {\bibinfo {author} {\bibfnamefont {J.~D.}\ \bibnamefont
  {Malcolm}}\ and\ \bibinfo {author} {\bibfnamefont {E.~J.}\ \bibnamefont
  {Nicol}},\ }\href {\doibase 10.1103/PhysRevB.90.035405} {\bibfield  {journal}
  {\bibinfo  {journal} {Phys. Rev. B}\ }\textbf {\bibinfo {volume} {90}},\
  \bibinfo {pages} {035405} (\bibinfo {year} {2014})}\BibitemShut {NoStop}%
\bibitem [{\citenamefont {Akrap}\ \emph {et~al.}(2016)\citenamefont {Akrap},
  \citenamefont {Hakl}, \citenamefont {Tchoumakov}, \citenamefont {Crassee},
  \citenamefont {Kuba}, \citenamefont {Goerbig}, \citenamefont {Homes},
  \citenamefont {Caha}, \citenamefont {Nov\'ak}, \citenamefont {Teppe},
  \citenamefont {Desrat}, \citenamefont {Koohpayeh}, \citenamefont {Wu},
  \citenamefont {Armitage}, \citenamefont {Nateprov}, \citenamefont
  {Arushanov}, \citenamefont {Gibson}, \citenamefont {Cava}, \citenamefont
  {van~der Marel}, \citenamefont {Piot}, \citenamefont {Faugeras},
  \citenamefont {Martinez}, \citenamefont {Potemski},\ and\ \citenamefont
  {Orlita}}]{Akrap2016}%
  \BibitemOpen
  \bibfield  {author} {\bibinfo {author} {\bibfnamefont {A.}~\bibnamefont
  {Akrap}}, \bibinfo {author} {\bibfnamefont {M.}~\bibnamefont {Hakl}},
  \bibinfo {author} {\bibfnamefont {S.}~\bibnamefont {Tchoumakov}}, \bibinfo
  {author} {\bibfnamefont {I.}~\bibnamefont {Crassee}}, \bibinfo {author}
  {\bibfnamefont {J.}~\bibnamefont {Kuba}}, \bibinfo {author} {\bibfnamefont
  {M.~O.}\ \bibnamefont {Goerbig}}, \bibinfo {author} {\bibfnamefont {C.~C.}\
  \bibnamefont {Homes}}, \bibinfo {author} {\bibfnamefont {O.}~\bibnamefont
  {Caha}}, \bibinfo {author} {\bibfnamefont {J.}~\bibnamefont {Nov\'ak}},
  \bibinfo {author} {\bibfnamefont {F.}~\bibnamefont {Teppe}}, \bibinfo
  {author} {\bibfnamefont {W.}~\bibnamefont {Desrat}}, \bibinfo {author}
  {\bibfnamefont {S.}~\bibnamefont {Koohpayeh}}, \bibinfo {author}
  {\bibfnamefont {L.}~\bibnamefont {Wu}}, \bibinfo {author} {\bibfnamefont
  {N.~P.}\ \bibnamefont {Armitage}}, \bibinfo {author} {\bibfnamefont
  {A.}~\bibnamefont {Nateprov}}, \bibinfo {author} {\bibfnamefont
  {E.}~\bibnamefont {Arushanov}}, \bibinfo {author} {\bibfnamefont {Q.~D.}\
  \bibnamefont {Gibson}}, \bibinfo {author} {\bibfnamefont {R.~J.}\
  \bibnamefont {Cava}}, \bibinfo {author} {\bibfnamefont {D.}~\bibnamefont
  {van~der Marel}}, \bibinfo {author} {\bibfnamefont {B.~A.}\ \bibnamefont
  {Piot}}, \bibinfo {author} {\bibfnamefont {C.}~\bibnamefont {Faugeras}},
  \bibinfo {author} {\bibfnamefont {G.}~\bibnamefont {Martinez}}, \bibinfo
  {author} {\bibfnamefont {M.}~\bibnamefont {Potemski}}, \ and\ \bibinfo
  {author} {\bibfnamefont {M.}~\bibnamefont {Orlita}},\ }\href {\doibase
  10.1103/PhysRevLett.117.136401} {\bibfield  {journal} {\bibinfo  {journal}
  {Phys. Rev. Lett.}\ }\textbf {\bibinfo {volume} {117}},\ \bibinfo {pages}
  {136401} (\bibinfo {year} {2016})}\BibitemShut {NoStop}%
\bibitem [{\citenamefont {Habibi}\ \emph {et~al.}(2022)\citenamefont {Habibi},
  \citenamefont {Musthofa}, \citenamefont {Adibi}, \citenamefont {Ekstr{\"o}m},
  \citenamefont {Schmidt},\ and\ \citenamefont {Hasdeo}}]{Habibi2022}%
  \BibitemOpen
  \bibfield  {author} {\bibinfo {author} {\bibfnamefont {A.}~\bibnamefont
  {Habibi}}, \bibinfo {author} {\bibfnamefont {A.~Z.}\ \bibnamefont
  {Musthofa}}, \bibinfo {author} {\bibfnamefont {E.}~\bibnamefont {Adibi}},
  \bibinfo {author} {\bibfnamefont {J.}~\bibnamefont {Ekstr{\"o}m}}, \bibinfo
  {author} {\bibfnamefont {T.~L.}\ \bibnamefont {Schmidt}}, \ and\ \bibinfo
  {author} {\bibfnamefont {E.~H.}\ \bibnamefont {Hasdeo}},\ }\href
  {https://dx.doi.org/10.1088/1367-2630/ac706d} {\bibfield  {journal} {\bibinfo
   {journal} {New J. Phys.}\ }\textbf {\bibinfo {volume} {24}},\ \bibinfo
  {pages} {063003} (\bibinfo {year} {2022})}\BibitemShut {NoStop}%
\bibitem [{\citenamefont {Liu}\ and\ \citenamefont {Dai}(2020)}]{J-Liu2020}%
  \BibitemOpen
  \bibfield  {author} {\bibinfo {author} {\bibfnamefont {J.}~\bibnamefont
  {Liu}}\ and\ \bibinfo {author} {\bibfnamefont {X.}~\bibnamefont {Dai}},\
  }\href {\doibase 10.1038/s41524-020-0299-4} {\bibfield  {journal} {\bibinfo
  {journal} {npj Comput. Mater.}\ }\textbf {\bibinfo {volume} {6}},\ \bibinfo
  {pages} {57} (\bibinfo {year} {2020})}\BibitemShut {NoStop}%
\bibitem [{\citenamefont {Ochoa}\ and\ \citenamefont
  {Asenjo-Garcia}(2020)}]{Ochoa2020}%
  \BibitemOpen
  \bibfield  {author} {\bibinfo {author} {\bibfnamefont {H.}~\bibnamefont
  {Ochoa}}\ and\ \bibinfo {author} {\bibfnamefont {A.}~\bibnamefont
  {Asenjo-Garcia}},\ }\href {\doibase 10.1103/PhysRevLett.125.037402}
  {\bibfield  {journal} {\bibinfo  {journal} {Phys. Rev. Lett.}\ }\textbf
  {\bibinfo {volume} {125}},\ \bibinfo {pages} {037402} (\bibinfo {year}
  {2020})}\BibitemShut {NoStop}%
\bibitem [{\citenamefont {Hillebrecht}\ \emph {et~al.}(1997)\citenamefont
  {Hillebrecht}, \citenamefont {Schmidt}, \citenamefont {Rotter}, \citenamefont
  {Thiele}, \citenamefont {Z{\"o}nnchen}, \citenamefont {Bengel}, \citenamefont
  {Cantow}, \citenamefont {Magonov},\ and\ \citenamefont
  {Whangbo}}]{H-Hillebrecht1997}%
  \BibitemOpen
  \bibfield  {author} {\bibinfo {author} {\bibfnamefont {H.}~\bibnamefont
  {Hillebrecht}}, \bibinfo {author} {\bibfnamefont {P.~J.}\ \bibnamefont
  {Schmidt}}, \bibinfo {author} {\bibfnamefont {H.~W.}\ \bibnamefont {Rotter}},
  \bibinfo {author} {\bibfnamefont {G.}~\bibnamefont {Thiele}}, \bibinfo
  {author} {\bibfnamefont {P.}~\bibnamefont {Z{\"o}nnchen}}, \bibinfo {author}
  {\bibfnamefont {H.}~\bibnamefont {Bengel}}, \bibinfo {author} {\bibfnamefont
  {H.~J.}\ \bibnamefont {Cantow}}, \bibinfo {author} {\bibfnamefont {S.~N.}\
  \bibnamefont {Magonov}}, \ and\ \bibinfo {author} {\bibfnamefont {M.~H.}\
  \bibnamefont {Whangbo}},\ }\href {\doibase
  https://doi.org/10.1016/S0925-8388(96)02465-6} {\bibfield  {journal}
  {\bibinfo  {journal} {J. Alloys Compd.}\ }\textbf {\bibinfo {volume} {246}},\
  \bibinfo {pages} {70} (\bibinfo {year} {1997})}\BibitemShut {NoStop}%
\bibitem [{Sup()}]{SuppMater}%
  \BibitemOpen
  \href@noop {} {\bibinfo  {journal} {See Supplemental Material at
  http://link.aps.org/xxx, which includes a detailed description of
  computational methods as well as supplemental figures and tables}\
  }\BibitemShut {NoStop}%
\bibitem [{\citenamefont {Zhang}\ \emph {et~al.}(2022)\citenamefont {Zhang},
  \citenamefont {Lin}, \citenamefont {Moreo}, \citenamefont {Maier},
  \citenamefont {Alvarez},\ and\ \citenamefont {Dagotto}}]{Y-Zhang2022}%
  \BibitemOpen
\bibfield  {journal} {  }\bibfield  {author} {\bibinfo {author} {\bibfnamefont
  {Y.}~\bibnamefont {Zhang}}, \bibinfo {author} {\bibfnamefont {L.-F.}\
  \bibnamefont {Lin}}, \bibinfo {author} {\bibfnamefont {A.}~\bibnamefont
  {Moreo}}, \bibinfo {author} {\bibfnamefont {T.~A.}\ \bibnamefont {Maier}},
  \bibinfo {author} {\bibfnamefont {G.}~\bibnamefont {Alvarez}}, \ and\
  \bibinfo {author} {\bibfnamefont {E.}~\bibnamefont {Dagotto}},\ }\href
  {\doibase 10.1103/PhysRevB.105.174410} {\bibfield  {journal} {\bibinfo
  {journal} {Phys. Rev. B}\ }\textbf {\bibinfo {volume} {105}},\ \bibinfo
  {pages} {174410} (\bibinfo {year} {2022})}\BibitemShut {NoStop}%
\bibitem [{\citenamefont {Bastos}\ \emph {et~al.}(2019)\citenamefont {Bastos},
  \citenamefont {Besse}, \citenamefont {Da~Silva},\ and\ \citenamefont
  {Sipahi}}]{CMO-Bastos2019}%
  \BibitemOpen
  \bibfield  {author} {\bibinfo {author} {\bibfnamefont {C.~M.~O.}\
  \bibnamefont {Bastos}}, \bibinfo {author} {\bibfnamefont {R.}~\bibnamefont
  {Besse}}, \bibinfo {author} {\bibfnamefont {J.~L.~F.}\ \bibnamefont
  {Da~Silva}}, \ and\ \bibinfo {author} {\bibfnamefont {G.~M.}\ \bibnamefont
  {Sipahi}},\ }\href {\doibase 10.1103/PhysRevMaterials.3.044002} {\bibfield
  {journal} {\bibinfo  {journal} {Phys. Rev. Mater.}\ }\textbf {\bibinfo
  {volume} {3}},\ \bibinfo {pages} {044002} (\bibinfo {year}
  {2019})}\BibitemShut {NoStop}%
\bibitem [{\citenamefont {Liu}\ \emph {et~al.}(2021)\citenamefont {Liu},
  \citenamefont {Chu}, \citenamefont {Zhang}, \citenamefont {Yu},\ and\
  \citenamefont {Yao}}]{G-Liu2021}%
  \BibitemOpen
  \bibfield  {author} {\bibinfo {author} {\bibfnamefont {G.-B.}\ \bibnamefont
  {Liu}}, \bibinfo {author} {\bibfnamefont {M.}~\bibnamefont {Chu}}, \bibinfo
  {author} {\bibfnamefont {Z.}~\bibnamefont {Zhang}}, \bibinfo {author}
  {\bibfnamefont {Z.-M.}\ \bibnamefont {Yu}}, \ and\ \bibinfo {author}
  {\bibfnamefont {Y.}~\bibnamefont {Yao}},\ }\href {\doibase
  https://doi.org/10.1016/j.cpc.2021.107993} {\bibfield  {journal} {\bibinfo
  {journal} {Comput. Phys. Commun.}\ }\textbf {\bibinfo {volume} {265}},\
  \bibinfo {pages} {107993} (\bibinfo {year} {2021})}\BibitemShut {NoStop}%
\bibitem [{\citenamefont {Liu}\ \emph {et~al.}(2023)\citenamefont {Liu},
  \citenamefont {Zhang}, \citenamefont {Yu},\ and\ \citenamefont
  {Yao}}]{G-Liu2022}%
  \BibitemOpen
  \bibfield  {author} {\bibinfo {author} {\bibfnamefont {G.-B.}\ \bibnamefont
  {Liu}}, \bibinfo {author} {\bibfnamefont {Z.}~\bibnamefont {Zhang}}, \bibinfo
  {author} {\bibfnamefont {Z.-M.}\ \bibnamefont {Yu}}, \ and\ \bibinfo {author}
  {\bibfnamefont {Y.}~\bibnamefont {Yao}},\ }\href {\doibase
  https://doi.org/10.1016/j.cpc.2023.108722} {\bibfield  {journal} {\bibinfo
  {journal} {Comput. Phys. Commun.}\ }\textbf {\bibinfo {volume} {288}},\
  \bibinfo {pages} {108722} (\bibinfo {year} {2023})}\BibitemShut {NoStop}%
\bibitem [{\citenamefont {see}()}]{refractiveindex}%
  \BibitemOpen
  \bibfield  {author} {\bibinfo {author} {\bibnamefont {see}},\ }\href
  {https://refractiveindex.info} {\bibinfo  {journal}
  {https://refractiveindex.info}\ }\BibitemShut {NoStop}%
\bibitem [{\citenamefont {Wang}\ \emph {et~al.}(2022)\citenamefont {Wang},
  \citenamefont {Jiang}, \citenamefont {Li},\ and\ \citenamefont
  {Xiao}}]{J-Wang2022}%
  \BibitemOpen
\bibfield  {journal} {  }\bibfield  {author} {\bibinfo {author} {\bibfnamefont
  {J.}~\bibnamefont {Wang}}, \bibinfo {author} {\bibfnamefont {C.}~\bibnamefont
  {Jiang}}, \bibinfo {author} {\bibfnamefont {W.}~\bibnamefont {Li}}, \ and\
  \bibinfo {author} {\bibfnamefont {X.}~\bibnamefont {Xiao}},\ }\href {\doibase
  https://doi.org/10.1002/adom.202102436} {\bibfield  {journal} {\bibinfo
  {journal} {Adv. Opt. Mater.}\ }\textbf {\bibinfo {volume} {10}},\ \bibinfo
  {pages} {2102436} (\bibinfo {year} {2022})}\BibitemShut {NoStop}%
\bibitem [{\citenamefont {Duan}\ \emph {et~al.}(2024)\citenamefont {Duan},
  \citenamefont {Ma}, \citenamefont {Zhang}, \citenamefont {Jiang},
  \citenamefont {Zhang}, \citenamefont {Cui}, \citenamefont {Yu},\ and\
  \citenamefont {Yao}}]{J-Duan2024}%
  \BibitemOpen
  \bibfield  {author} {\bibinfo {author} {\bibfnamefont {J.}~\bibnamefont
  {Duan}}, \bibinfo {author} {\bibfnamefont {D.-S.}\ \bibnamefont {Ma}},
  \bibinfo {author} {\bibfnamefont {R.-W.}\ \bibnamefont {Zhang}}, \bibinfo
  {author} {\bibfnamefont {W.}~\bibnamefont {Jiang}}, \bibinfo {author}
  {\bibfnamefont {Z.}~\bibnamefont {Zhang}}, \bibinfo {author} {\bibfnamefont
  {C.}~\bibnamefont {Cui}}, \bibinfo {author} {\bibfnamefont {Z.-M.}\
  \bibnamefont {Yu}}, \ and\ \bibinfo {author} {\bibfnamefont {Y.}~\bibnamefont
  {Yao}},\ }\href {\doibase https://doi.org/10.1002/adfm.202313067} {\bibfield
  {journal} {\bibinfo  {journal} {Adv. Funct. Mater.}\ }\textbf {\bibinfo
  {volume} {34}},\ \bibinfo {pages} {2313067} (\bibinfo {year}
  {2024})}\BibitemShut {NoStop}%
\end{thebibliography}%

\bibliographystyle{apsrev4-1.bst}

\end{document}